\documentclass[%
preprint,
 amsmath,amssymb,
pra,
]{revtex4-1}

\usepackage{graphicx}
\graphicspath{{doc/graphics/}}
\usepackage{dcolumn}
\usepackage{bm}
\usepackage{xfrac}

\newcommand{\bra}[1]{\langle #1|}
\newcommand{\ket}[1]{|#1\rangle}

\newcommand{\comment}[1]{}

\newcommand{\e}[1]{\ensuremath{\times 10^{#1}}}
\newcommand{\OHgT}[0]{380$\, \mathrm{K}$}
\newcommand{\OHgS}[0]{$1.6\e{-15}$  $\left[\sqrt{\mathrm{Hz}^{-1}}\right]$}
\newcommand\T{\rule{0pt}{2.6ex}}       
\newcommand{\mcS}[0]{$\mathcal{S}\,$}

\makeatletter
\newcommand\xleftrightarrow[2][]{%
  \ext@arrow 9999{\longleftrightarrowfill@}{#1}{#2}}
\newcommand\longleftrightarrowfill@{%
  \arrowfill@\leftarrow\relbar\rightarrow}
\makeatother

\begin{document}


\title{A Two-Photon E1-M1 Optical Clock}



\author{E. A. Alden}
\email{ealden@umich.edu}
\author{K.R. Moore}
\email{kaimoore@umich.edu}
\author{A. E. Leanhardt}
\email{aehardt@umich.edu}
\homepage{http://www.umich.edu/\textasciitilde aehardt/}
\affiliation{Departments of Physics and Applied Physics, University of Michigan, Ann Arbor, Michigan 48109-1040, USA}


\date{\today}

\begin{abstract}
  An allowed E1-M1 excitation scheme creates optical access to the ${^1S_0} \rightarrow {^3P_0}$ clock transition in group II type atoms. This method does not require the hyperfine mixing or application of an external magnetic field of other optical clock systems. The advantages of this technique include a Doppler-free excitation scheme and increased portability with the use of vapor cells. We will discuss technical mechanisms of a monochromatic excitation scheme for a hot E1-M1 clock and briefly discuss a bichromatic scheme to eliminate light shifts. We determine the optimal experimental parameters for Hg, Yb, Ra, Sr, Ba, Ca,  Mg, and Be and calculate that neutral Hg has ideal properties for a hot, portable frequency standard.
\end{abstract}


\maketitle

The rapid advancement in optical frequency standards has seen three different systems hold the mantle of best stability, $\mathcal{S}$, in the past year \cite{Chou2010,Hinkley2013,Bloom2014}. Increases in accuracy of optical clocks advance popular technologies such as global positioning systems (GPS), permit testing of fundamental physics constants, and have the potential to make local measurements of the gravitational redshift \cite{Chou2010a}. A map of the Earth's geodesy measured with the precision of an optical frequency standard will require a mobile atomic clock. Such an optical frequency standard can be achieved in a portable vapor cell by addressing the atoms with a two-photon, Doppler-free spectroscopy scheme. \

An E1-M1 optical clock is a frequency standard based on a two-photon excitation from the ground state to the clock state by a pair of electric (E1) and magnetic (M1) dipole allowed transitions. We characterize the atom-light interactions and determine the optimal experimental parameters for a selection of group II type atoms. We present the ideal temperature and laser beam radius to maximize the E1-M1 excitation rate and to optimize the E1-M1 optical clock stability. We find that neutral Hg is the ideal atomic system for an E1-M1 optical clock and discuss in detail Hg specific parameters. 

The benchmark by which frequency standards are compared is the stability $\mathcal{S}$, which is the rate at which minimum instability $\sigma_\nu$ can be attained where $\sigma_\nu(\tau) = \mathcal{S}/ \sqrt{\tau}$ and $\tau$ is the total measurement time. The stability has units of $\left[\sqrt{\mathrm{Hz}^{-1}}\right]$ and is fundamentally limited by\

\begin{equation}
\mathcal{S} = \frac{\Delta\nu}{\nu}\sqrt{\frac{\mathcal{T}}{N_D}} 
\label{eq:stab}
\end{equation}

\noindent where $\nu$ is the fundamental frequency of the standard, $\Delta\nu$ is the effective linewidth of the transition, $\mathcal{T}$ is the period of each detection cycle, and $N_D$ is the effective number of atoms that are detected each experiment period.

\begin{figure}
\begin{center}
\includegraphics[width=2.75in]{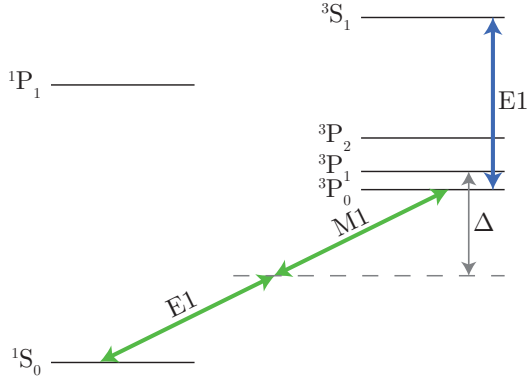}
\caption{\textbf{Two-Photon Clock Level Structure} This is the prototypical, optical clock level structure with a ${^3P_0}$ clock state. The electric field of one photon and the magnetic field of a second, degenerate photon directly couple the ${^1S_0}$ ground state to the clock state by coupling to the intermediate ${^3P_1}$ level with some detuning $\Delta$. A sample detection channel for a hot clock is the ${^3P_0} \xleftrightarrow{\text{E1}} {^3S_1}$, E1 allowed transition.
\label{Levels}}
\end{center}
\end{figure}

 A specific advantage of the hot E1-M1 optical clock compared with the other optical frequency standards is a large increase in $N_D$ and only a small increase in linewidth $\Delta\nu$ due to the thermal environment. The current generation of precision atomic clocks require extensive state preparation to remove first-order Doppler effects as the largest broadening mechanism, a process that limits the number of atoms that can be addressed. The estimated stability of a hot (\OHgT) Hg E1-M1 clock with the experimental constraints listed in section \ref{tempw} is \OHgS. The simplicity of the E1-M1 system will increase portability for metrology applications compared with current cold atomic clocks.

\section{${^1S_0}\rightarrow{^3P_0}$ e1-M1 Transition}
State-of-the-art optical clocks utilize the ${^1S_0} \rightarrow {^3P_0}$ clock transition. This is because relaxation of the ${^3P_0}$ clock level to the ${^1S_0}$ ground level is forbidden for all single-photon electric or magnetic radiation types. Forbidden relaxation channels create the narrow-linewidths required for precise time-keeping in atomic systems. However, all current clock systems operate by mixing the clock level with a nearby level ($^3P_1$) that does have weakly-allowed $E1$-coupling to the ground state. Mixing of ${}^3P_0$ and ${}^3P_1$ can be done by selecting isotopes with hyperfine structure \cite{Chou2010} or by applying a magnetic field \cite{Barber2006}. While mixing the levels is necessary to make them electric dipole coupled, it also reduces the lifetime. This can potentially limit the ultimate precision of the clock.\

Figure \ref{Levels} shows an alternative excitation scheme with the allowed E1-M1 two-photon transition along ${^1S_0} \xleftrightarrow{\text{E1}} {^3P_1} \xleftrightarrow{\text{M1}} {^3P_0}$ \cite{*[{The ${^1S_0} \xleftrightarrow{\text{E1}} {^3P_1}$ transition is allowed by virtue of spin-orbit coupling between the ${^3P_1}$ and ${^1P_1}$ levels. For an example in Hg see:  }][{}] Mcconnell1969}. This coupling has been observed previously in highly charged ions \cite{Schaffer1999,Karasiev1992}. As an allowed transition it provides optical access to all isotopes of group II type atoms. This excitation can be implemented using counter-propagating photons with either degenerate frequencies to eliminate first-order Doppler broadening (Sections \ref{mono},\ref{mono2}) or with non-degenerate frequencies chosen to offset light shifts (Section \ref{bich}). \

\begin{figure}
\begin{center}
\includegraphics[width=2.75in]{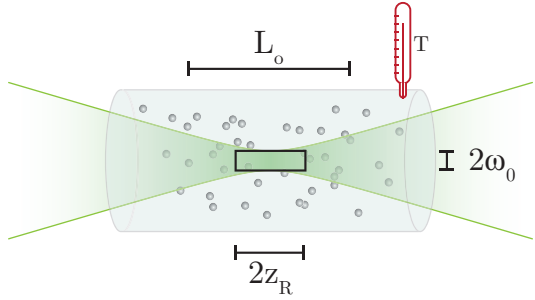}
\caption{\textbf{Hot Optical Clock} This diagram of a monochromatic laser in a vapor cell depicts the experimental system. The detection length $L_o$ is the detection optics aperture and the laser beam radius is $\omega_0$. The Rayleigh range, $z_R$, that limits the interrogation region is shown in this graphic. The area enclosed by a box is the Rayleigh-limited interrogation region of the atoms. 
\label{Vapor}}
\end{center}
\end{figure}

Figure \ref{Vapor} shows a typical experimental setup for a hot optical clock. The viability of a hot vapor cell clock will depend on the effective rate of detected atoms: 

\begin{eqnarray}
\dot{N}_{D} = P_{D} \times \dot{N}_{^3P_0}
\label{eq:ND}
\end{eqnarray}

\noindent where $P_{D}$ is the probability of detecting an atom in the $^3P_0$ clock-level, and $\dot{N}_{^3P_0}$ is the effective rate of atoms excited to the clock level. The two primary experimental parameters which require optimization are vapor cell temperature $T$ and laser beam radius $\omega_0$. The effective rate of atoms excited to the $^3P_0$-level is given by:

\begin{eqnarray}
\dot{N}_{^3P_0} = P_{^3P_0}(T,\omega_0) \times \dot{N}_{tot}(T,\omega_0)
\label{eq:NPPNt}
\end{eqnarray}

\noindent where $P_{^3P_0}$ is the probability a single atom in the excitation region has been excited to the $^3P_0$ level and $\dot{N}_{tot}$ is the rate of atoms flowing through the interrogation region. In a thermal environment the interrogation time of an atom by the excitation laser is always much less than the time required to coherently transfer the full population to the excited state; there is no risk of Rabi flopping. An increase in laser power is therefore always beneficial because it increases the two-photon Rabi frequency and by extension the probability of excitation the the $^3P_0$-level in a time-limited measurement. The temperature and laser beam radius contribution to overall rates and stability will be explained in Section \ref{DynP}.

\subsection{Two-Photon Rabi Frequency}

For the purposes of this paper, we will consider a system where the atom is excited with a single laser (monochromatic) or pair of lasers (bichromatic) whose frequencies are far off resonance from the allowed E1 or M1 transitions. To satisfy the selection rules of the transition, the electric field vector of one excitation photon must be parallel to the magnetic field vector of the other excitation photon. This alignment can be realized utilizing either the Lin $\perp$ Lin or $\sigma^+/\sigma^-$ polarization scheme described by Dalibard and Cohen-Tannoudji \cite{Dalibard1989}. These schemes satisfy the selection rules and ensure that any clock excitation is the product of excitation from counter-propagating beams and thus reduces or eliminates first-order Doppler-broadening.\

 Our proposed system satisfies the constraints of adiabatic elimination \cite{Shore1990, Brion2007}, specifically $\Delta >> \Omega_1,\Omega_2,\delta $ where $\Omega_i$ is the two-level Rabi frequency of each E1 and M1 transition, $\delta$ is the two-photon detuning from the unperturbed transition frequency, and $\Delta$ is the minimum detuning of an excitation photon's energy from the intermediate $^3P_1$ level, see Fig. \ref{Levels}. In this limit, the two-photon Rabi frequency for an atom addressed by a pair of photons, where $\delta$ is chosen to offset the light shift, is given by \cite{Moler1992} :

\begin{eqnarray}
\Omega_{R 2\gamma} = \frac{2I}{\hbar^2 c^2 \epsilon_0}\frac{\bra{^3P_0}|\mu| \ket{^3P_1}_{M1}\bra{^3P_1}|D| \ket{^1S_0}_{E1}}{\Delta}
\label{eq:2gRates}
\end{eqnarray}

\noindent where $I$ is the peak intensity of the excitation laser, $\bra{^3P_0}|\mu| \ket{^3P_1}_{M1}$ is the reduced matrix element for the magnetic dipole (M1) transition, $\bra{^3P_1}|D| \ket{^1S_0}_{E1}$ is the reduced matrix element for the electric dipole (E1) transition. \

\begin{table}%
\caption{Reduced matrix elements for the electric-dipole $\bra{nsnp {^3P}_1}|\textbf{D}|\ket{n{s^2}\, {^1S_0}}$ intercombination transition (E1) and the magnetic-dipole $\bra{nsnp {^3P}_0}|\boldsymbol{\mu}|\ket{nsnp {^3P_1}}$ transition (M1) for each candidate element. Matrix element values are in a.u. For monochromatic excitation, the two-photon Rabi frequency, $\Omega_{R 2\gamma}$, is shown for unit intensity (1 W/m$^2$). A prototypical intensity for this scheme is $6\e{6} [W/m^2]$.} 
\label{tgR}\centering %
\begin{tabular}{ c | cccc }
Atom &  n &E1/$ea_{_0}$ & M1/$\mu_{_B}$ & $\Omega_{R 2\gamma}/I [Hz]$ \\
\hline
Ra &  $  7$ & $1.2$ \cite{Bieron2007} & $\sqrt{2}$ \cite{Curtis2001a} & $7.1\e{-5}$ \T \\
Ba &  $  6$ & $0.45$ \cite{Dzuba2000} & $\sqrt{2}$  \cite{Curtis2001a}  & $3\e{-5}$ \\
Yb &  $  6$ & $0.54$ \cite{Beloy2012} & $\sqrt{2}$ \cite{Beloy2012} & $2.5\e{-5}$ \\
Hg &  $  6$ & $0.44$ \cite{NISTASD} & $\sqrt{2}$  \cite{Curtis2001a}  & $9.3\e{-6}$ \\
Sr &  $  5$ & $0.15$ \cite{Porsev2001} & $\sqrt{2}$  \cite{Curtis2001a}  & $8.8\e{-6}$ \\
Ca &  $  4$ & $0.036$ \cite{Porsev2001} & $\sqrt{2}$  \cite{Curtis2001a}  & $2\e{-6}$ \\
Mg &  $  3$ & $0.0057$ \cite{Porsev2001} & $\sqrt{2}$ \cite{NISTASD} & $2.2\e{-7}$ \\
Be &  $  2$ & $0.00024$ \cite{NISTASD} & $\sqrt{2}$  \cite{Curtis2001a}  & $9.3\e{-9}$ 
\end{tabular}
\end{table}

The E1-M1 coupling will also occur via the $^1P_1$ intermediate level. In the case of Hg it will constitute as much as $37 \%$ of the Rabi frequency, where its contribution is maximum for the degenerate excitation scheme. We omit this favorable contribution from the rate and stability simulations for simplicity, but experiments can anticipate an enhancement. Estimated and observed electric- and magnetic-dipole matrix elements are shown for the group II type atoms in Table \ref{tgR}. We also provide the estimated two-photon Rabi frequency for the degenerate photon case with the experimental parameters defined in Table \ref{EAss}.

\subsection{Probability of Clock Excitation}
The effective excitation rate ($\dot{N}_{^3P_0}$) for an atomic vapor with static experimental settings depends on the probability of exciting an atom to the clock level ($P_{^3P_0}$) and the atomic interrogation rate ($\dot{N}_{tot}$). 

With adiabatic elimination of the intermediate level, the probability of exciting the atom into the $^3P_0$ clock level follows \cite{Shore1990,Brion2007}:

\begin{equation}
P_{^3P_0} = \frac{\Omega_{R 2\gamma}^2}{\Omega_{R 2\gamma}^2+\delta^2}\sin^2\left(\frac{\sqrt{\Omega_{R 2\gamma}^2+\delta^2}}{2}\bar{t}\right)
\label{eq:probclock}
\end{equation}

\noindent where $\bar{t}$ is the average interrogation time of the atoms and $\delta$ is the detuning from the light-shifted resonant frequency. \

For simplicity, excitation probability is calculated by assuming that the dominant broadening mechanism introduces an effective, constant detuning $\delta$. The ultimate excitation rate $\dot{N}_{^3P_0}$ for this approximation typically varies from a more exact calculation by less than a factor of ten. The effective broadening of a hot E1-M1 clock is much larger than the two-photon Rabi frequency due to large first-order Doppler broadening: $\Delta \nu_{D1}>>\Omega_{R 2\gamma}$ so the excitation probability, $P_{^3P_0}$, has two characteristic regimes. There is a time-limited regime, where transit-time (time-of-flight) broadening, $\Delta\nu_{_{TT}} =\sfrac{1}{\bar{t}}$, is the dominant scaling feature of the excitation probability. The time-limited probability of excitation scales quadratically in time as

\begin{equation}
P_{^3P_0} \approx \left(\frac{\Omega_{R 2\gamma}\bar{t}_B}{2}\right)^2.
\label{eq:quadcl}
\end{equation}

 In a velocity-limited regime where $\Delta \nu_{D1} >> \Delta\nu_{_{TT}} $ the effective detuning $\delta$ is approximately the first-order Doppler broadening $\Delta \nu_{D1}$. Here the probability of clock excitation resembles a saturated system and can be simplified as:

\begin{equation}
P_{^3P_0}(T,\omega_0) \approx \frac{\Omega_{R 2\gamma}^2}{\Delta \nu_{D1}^2}.
\label{eq:satclock}
\end{equation}


\section{monochromatic  E1-M1 clock\newline
experimental parameters}
\label{mono}
\subsection{Static Parameters}
\label{tempw}

Most experimental parameters can be optimized independently. Table \ref{EAss} lists the static magnitudes we will assume for these independent experimental parameters. The magnitudes were conservatively selected to match existing or easily attainable levels. Importantly, these experimental magnitudes can be constructed in a portable package which will permit mobile, optical frequency measurements.

\begin{table} [b]%
\caption{Experimental parameters used for the simulation of a hot, vapor cell E1-M1 optical clock.} 
\label{EAss}\centering %
\begin{tabular*}{3.45in}{l c @{\extracolsep{\fill}} r}
Parameter& & Value \\
\hline
Power (C.W.) & & 10 W\T\\
Laser Linewidth && 1 kHz\\
Retroreflection Misalignment $\theta$    & & 0.1 milliradian\\
Photon Collection Efficiency $P_{pc}$& & 1 \%\\
Optical Damage Threshold& & 800 K\\
Temperature Instability $\sigma_{\bar{T}}$& &0.1 K\\
Optical Detection Length $L_o$ & & 2 cm\\
Experiment Period $\mathcal{T}$&& 1 s
\end{tabular*}
\end{table}

We estimate that narrow-linewidth power can be achieved at approximately the 10 W level. We have generated 8 W of portable, narrow-linewidth light for a monochromatic Hg experiment in a preliminary device and estimate that more power is possible while preserving this portability \cite{Alden2014}. The power levels discussed here will contribute a sizable light shift to the system that can be effectively quantified. Section \ref{LSStab} discusses the light shift in detail.

All group II type atoms, with the exception of Hg, improve in overall excitation rate and stability with increased temperature because atom-atom interactions are minimal. It was therefore necessary to impose in our calculations a temperature ceiling of 800 K as a damage threshold for the physical optics and detectors in an experimental apparatus. 

The size of the photon detector and imaging optics will limit the volume of atoms which can be observed. Since detectors exist with 2-cm diameter,  we anticipate that with a 1:1 imaging system a detection length, $L_o$, of 2-cm can be implemented. The detection length is shown in Figure \ref{Vapor} in relation to other experimental geometry parameters. 

The misalignment of the counterpropagating laser beams by angle $\theta$ is anticipated.  We impose a 0.1 milliradian limit in our simulation which has been previously realized in atom interferometry experiments \cite{Gustavson1997}. Increased precision of alignment will improve both the rates and the stability. We do not include a threshold for polarization rotation errors although such experimental features will degrade the rates and stability of the experiment.

An optical frequency standard will limit itself to a single atomic isotope. The natural abundance of the excluded isotopes will attenuate the excitation rates we calculate. Since isotope abundance is idiosyncratic to each atomic species, our calculations do not include the reduction to the number density that will be present in an experiment.

\subsection{Dynamic Parameters}
\label{DynP}

Adjustments in vapor cell temperature and laser beam radius can increase the number of atoms in the clock state or the precision of the standard for each atom species. Using thermal atoms instead of ultracold atoms increases the density and interrogation rate, a statistical advantage for this optical frequency standard. A disadvantage for thermal atoms is atomic-mass-dependent thermal speed which both limits interrogation to the transit time $\bar{t}$ of the atom and introduces sensitivity to Doppler effects ($\Delta\nu_{D1}$ and $\Delta\nu_{D2}$). Unencumbered motion also leads to non-zero atom-atom collision probability.

Excitation probability scales with the interrogation time of the atoms and is sensitive to laser beam radius and vapor cell temperature. We define the interrogation time $\bar{t}$ as the average time atoms spend passing through a volume of the laser beam enclosed by the Gaussian radius of the laser beam $\omega_0$ and an experimentally limited length of the laser beam path shown in Figure \ref{Vapor}. From calculations we find that the beam diameter is a reasonable approximation of the average distance $\bar{l}$ an atom travels through the interrogation region. Using the mean thermal velocity of an atom in the vapor cell $\bar{v}$ we can find the average interrogation time for the atoms which scales with temperature and laser beam radius as

\begin{eqnarray}
\bar{t} = \frac{\bar{l}}{\bar{v}} = 2\omega_0 \times \sqrt{\frac{\pi m}{8k_BT}}
\label{eq:tB}
\end{eqnarray}

\noindent where $k_B$ is Boltzmann's constant and $m$ is the mass of the atom. Mean interrogation time scales with the square root of atomic mass, so heavier atoms enjoy longer interrogation times. The mass of each candidate atom is shown in Table \ref{atomchar}\

\begin{table}[h]%
\caption{Mass and natural optical transition frequency $\nu_0$ for each candidate element. The specific wavelength for the monochromatic excitation scheme, $\lambda_{2\gamma}$, is also listed. Citations are included when the clock transition has been experimentally observed.} 
\label{atomchar}\centering %
\begin{tabular}{ c c | ccc }
Atom && m [amu]  & $\nu_0$ [Hz] & $\lambda_{2\gamma}$ [nm]\\
\hline
Hg &\cite{McFerran2012} & $200.6$ & $ 1.1\e{15}$   & 531 \T \\
Be && $9.1$ & $ 6.6\e{14}$  & 910\\
Mg && $24.3$ & $6.6\e{14}$   &915\\
Yb& \cite{Hong2005}& $173.1$ & $5.2\e{14}$  &1157\\
Ca&& $40.1$ & $4.5\e{14}$  &1319\\
Sr &\cite{Ludlow2006}& $87.6$ & $ 4.3\e{14}$ & 1397 \\
Ra && $226.0$ & $3.9\e{14}$ &1529 \\
Ba && $137.3$ & $ 3.7\e{14}$  & 1631
\end{tabular}
\end{table}

For the degenerate excitation case, mirror misalignment introduces first-order Doppler broadening,  $\Delta \nu_{D1}$, to atoms with velocity components co-linear with the laser beams. This is the second-largest broadening after transit broadening, and we use the characteristic width of this Doppler effect as a constant detuning $\delta$ in (\ref{eq:probclock}). We determine this harmful broadening from

\begin{eqnarray}
\Delta \nu_{D1}(T ,\theta) = k\bar{v}(T) \sin(\theta) 
\label{eq:1D}
\end{eqnarray}

\noindent  where $k$ is the wavenumber. The specific detuning, $\Delta \nu_{D1}$, for each atom with respect to vapor cell temperature and misalignment is listed in Table \ref{D1}. 
Typical experimental parameters (Table \ref{EAss}) introduce a broadening of 10-100 kHz for the candidate atoms. This can be compared to the estimated natural linewidth of 0.45 Hz for neutral Hg \cite{Mishra2001}.

\begin{table}[h]%
\caption{1st-order Doppler broadening introduced by mirror misalignment $\theta$. The milliradian misalignment threshold make the small angle approximation valid.} 
\label{D1}\centering %
\begin{tabular}{ c c }
Atom & $\Delta \nu_{D1}$($T$ [K],$\theta $[rad]) [Hz]   \\
\hline
Ra & $6.3\e{6}\sqrt{T} \times \theta $ \T \\
Ba & $7.6\e{6}\sqrt{T} \times \theta $\\
Yb & $9.6\e{6}\sqrt{T} \times \theta $\\
Sr & $1.1\e{7}\sqrt{T} \times \theta $\\
Ca & $1.7\e{7}\sqrt{T} \times \theta $\\
Hg & $1.9\e{7}\sqrt{T} \times \theta $\\
Mg & $3.2\e{7}\sqrt{T} \times \theta $\\
Be & $5.3\e{7}\sqrt{T} \times \theta $\\
\end{tabular}
\end{table}

The atom interrogation rate, $\dot{N}_{tot}$, depends on the temperature and interrogation volume. The interrogation volume is a region of high laser intensity limited either by the Rayleigh range or the detection optics. Figure \ref{Vapor} depicts these lengths in a vapor cell. When the volume is limited by detection optics, $V_{L_o}$, the volume of addressed atoms is the laser beam area multiplied by the detection length, $L_o$

\begin{eqnarray}
V_{L_o} &=& L_{o} \times \pi\omega_0^2\\
V_{L_o}(\omega_0) &\propto& \omega_0^2.
\label{eq:VLo}
\end{eqnarray}

\noindent In the Rayleigh range limit the volume, $V_{2z_R}$, is given by:

\begin{eqnarray}
V_{2z_R} &=& \frac{2\pi\omega_0^2}{\lambda} \times \pi\omega_0^2\\
V_{2z_R}(\omega_0) &\propto& \omega_0^4.
\label{eq:VLo}
\end{eqnarray}

\noindent From these equations, we can see that a large laser beam radius maximizes the volume and the system favors utilizing the full Rayleigh range as long as $2z_R \leq L_o$. \

\begin{figure}[!h]
\includegraphics[width=3.35in]{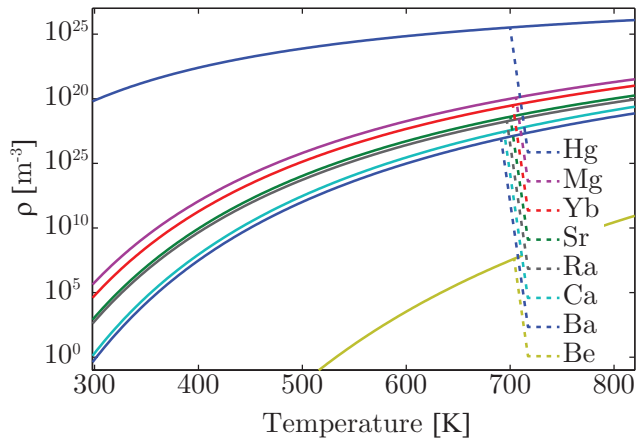}
\caption{\textbf{Vapor Cell Density} Number density of group II type atoms with respect to vapor cell temperature. The high number density of Hg gives it a statistical advantage for optical stability.  
\label{RhoT}}
\end{figure}

The number density of atoms, $\rho$, in the vapor cell can be calculated from published vapor pressure curves \cite{*[{The number density of Ra in this temperature range is not known, we extrapolate the curve for Ra from known high temperature values and assumed similarity to other group II type atoms .  }][{}] Alcock1984,Huber2006} and is shown in Figure \ref{RhoT}. The number density $\rho$ of all group II type atoms scales exponentially with temperature and can be expressed as

\begin{eqnarray}
\rho(T [K]) \propto e^{\frac{-10^4}{T}}.
\label{eq:RhT}
\end{eqnarray}

The rate of atom interrogation, $\dot{N}_{tot}$, is the product of the excitation volume, $V$, the number density at room temperature, $\rho$, and the rate at which atoms refresh in that volume. This refresh rate is the inverse of the average interrogation time $\bar{t}$. We preserve in curly brackets the different scaling behavior of detection- or Rayleigh-limited regimes. The interrogation rate is as follows:

\begin{eqnarray}
\dot{N}_{tot} &=& V\times\frac{\rho}{\bar{t}}\nonumber \\
 &=& {V_{L_o} \brace V_{2z_R}}\times\frac{\rho}{\bar{t}}\\
\label{eq:cbgeo}
\dot{N}_{tot}(T [K],\omega_0 [m])& \propto& {\omega_0^2 \brace \omega_0^4}\times\frac{\sqrt{T}}{\omega_0} e^{\frac{-10^4}{T}}
\label{eq:ntot}
\end{eqnarray}

\subsection{Probability of Clock Excitation}
The time-limited probability of clock excitation (\ref{eq:quadcl}) in a hot, vapor cell can be written explicitly in terms of the vapor cell temperature and laser beam radius:

\begin{equation}
P_{^3P_0}(T,\omega_0) \approx \Omega_{R 2\gamma}(\omega_0)^2\bar{t}(T,\omega_0)^2 \propto \frac{1}{T\omega_0^2}.
\label{eq:pclSIMP}
\end{equation}

\noindent This approximation provides less than 1\% disagreement with the $P_{^3P_0}$ scaling behavior for sub-millimeter laser beam radii in simulation.\

The Doppler-limited probability of excitation (\ref{eq:satclock}) can also be written explicitly in terms of the vapor cell temperature and laser beam radius:

\begin{equation}
P_{^3P_0}(T,\omega_0) \approx \frac{\Omega_{R 2\gamma}(\omega_0)^2}{\Delta \nu_{D1}(T)^2} \propto \frac{1}{T\omega_0^4}.
\label{eq:pclSIMP2}
\end{equation}

\noindent In both regimes a high excitation probability $P_{^3P_0}$ favors small laser beam radius and low temperature. Small laser beam radius is an intuitive advantage here because atoms are not interrogated long enough to undergo coherent Rabi flopping so a more intense laser will enhances the two-photon Rabi frequency. Lower temperatures lead to longer interrogation times, which increase the probability of exciting a single atom.

\subsection{Clock Excitation Rate}

The experimental parameters to maximize the excitation rate $\dot{N}_{^3P_0}$ are different than those for optimum stability. However, maximum $\dot{N}_{^3P_0}$ will be experimentally convenient to quantify and optimize broadening and rate parameters for the ultimate optical frequency standard.

The effective clock excitation rate $\dot{N}_{^3P_0}$ (\ref{eq:NPPNt}) based on the time-limited probability of excitation $P_{^3P_0}$ (\ref{eq:pclSIMP}) and the interrogation rate $\dot{N}_{tot}$ defined by (\ref{eq:ntot}) can now be reported in terms of temperature and laser beam radius:

\begin{eqnarray}
\dot{N}_{^3P_0} &=& P_{^3P_0} \times \dot{N}_{tot}\\
\dot{N}_{^3P_0}(T,\omega_0) &\propto& \frac{1}{T\omega_0^2} \times {\omega_0 \brace \omega_0^3}\sqrt{T} e^{\frac{-10^4}{T}}\\
 &\propto& \frac{e^{\frac{-10^4}{T}}}{\sqrt{T}}{\frac{1}{\omega_0} \brace \omega_0}
\label{eq:NTw}
\end{eqnarray}

\begin{figure}
\begin{center}
\includegraphics[width=2.75in]{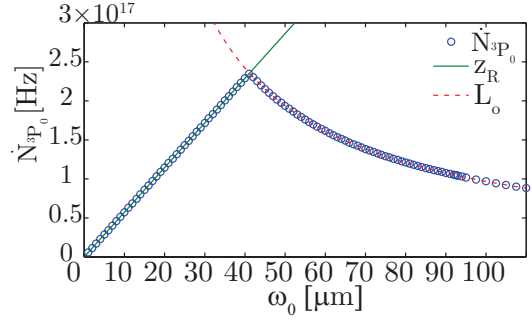}
\caption{\textbf{Optimal Laser Beam Size} Plot of effective excitation rate, $\dot{N}_{^3P_0}$, with respect to beam waist size, Eqn. \ref{eq:NPPNt} for neutral Hg at 448 K. Displays the asymptotic behavior in Rayleigh-limited and detection-limited regimes where the peak rates are found when the Rayleigh range $z_R$ is set to match the detection length $L_o$. In Hg this happens at 41 $\mu$m.  
\label{Tot}}
\end{center}
\end{figure}

\noindent The curly brackets continue to denote the detection- and Rayleigh-limited regimes of the experiment ${L_o \brace 2z_R}$. The optimal beam radius is the length where the Rayleigh range, $2z_R$, matches the detection length, $L_o$. For small $\omega_0$ where $2z_r<<L_o$, the overall rate increases with beam radius until $2z_R = L_o$, at which point the volume is detection limited and detection rates begin to decrease with $\omega_0$, as shown in Figure \ref{Tot}.

The calculated excitation rates for the group II type atoms, with optimal temperature and laser beam radius, are reported in Table \ref{Ntb} for the prototypical experimental settings listed in Table \ref{EAss}.

\subsection{Experimental Detection of $^3P_0$ Atoms }
The detection of moving atoms in the ${^3P_0}$ clock level will require a subsequent transition from the clock level to an electric-dipole-coupled level. In neutral Hg, the $6s6p {^3P_0}$ clock level is $E1$-coupled to a $6s7s {^3S_1}$ level by a $405$ nm photon. The $6s7s {^3S_1}$ rapidly cascades to the ground level, predominantly through the ${6s6s\,^3P_1}$ intermediate level \cite{Benck1989}. This cascade channel radiates at $436$ nm, a wavelength distinct from all others in the system.

The probability of detecting clock level occupation further depends on the collection efficiency of the experiment and the loss due to collision. Collisions are treated as a loss channel with a rate proportional to the interrogation length, $\bar{l} = 2\omega_0$, and the mean free path of the particles, $(\rho\sigma)^{-1}$, where $\rho$ is the number density of the atoms $[m^{-3}]$ and $\sigma$ is the Van der Waals radius $[m^2]$. The number density, $\rho$, exponentially increases with temperature (\ref{eq:RhT}). The risk of collision increases with interrogation length and cell temperature. The probability of no collision $P_{nc}$ is given by

\begin{eqnarray}
P_{nc}(T,\omega_0) &=& e^{-\rho(T)\sigma\bar{l}(\omega_0)}\\
P_{nc}(T [K],\omega_0 [m]) &\propto &e^{-\omega_0e^{\frac{-10^4}{T}}}
\label{eq:nocoll}
\end{eqnarray}

\noindent and primarily depends on the number density of atoms (see Figure \ref{RhoT}). Hg is the only atom with a non-negligible probability of collision for the temperatures and interrogation lengths of this calculation. This is due to its considerably higher density compared with the other species, but the higher collision rate does not eliminate its viability as an optical clock. See Section \ref{collshift} for details. 
Detection probability also depends on the photon collection efficiency of the imaging system. We assume a collection efficiency, $P_{pc}$, of 1\%, as listed in Table \ref{EAss}.\ The final probability of detection, $P_{D}$, is

\begin{equation}
P_D = P_{pc}\times P_{nc}.
\label{eq:PD}
\end{equation}

\subsection{Optimal Parameters for Detection Rate $\dot{N}_D$}
Table \ref{Ntb} reports the maximum calculated detection rates $\dot{N}_D$ for the hot monochromatic E1-M1 scheme. These results are presented for the group II type atoms Hg, Sr, Yb, Ca, Mg, Be, Ra, and Ba. Optimal temperatures $T$ and laser-beam radii $\omega_0$ are listed for each atomic species. The optimal temperature for Hg differs from the other elements due to collision rate sensitivity.

\begin{table} [htp]%
\caption{The detection rate of clock atoms, $\dot{N}_{D}$, at the optimal vapor cell temperature $T$ and laser beam radius $\omega_0$ for each group II type atom. The optimal $\omega_0$ in all cases is the radius where the Rayleigh range matches the detection length, $L_o$, of the detection optics which maximizes $\dot{N}_D$. These are not the values for optimal clock stability. } 
\label{Ntb}\centering %
\begin{tabular}{ c c c l}
Atom & $T$ [K] & $\omega_0 [\mu m]$ & $\,\,\, \dot{N}_{D}$ [$s^{-1}$]  \\
\hline
Hg & 448  & 41.0 & $1.8\e{+12}$ \T \\
Yb & 800  & 60.0 & $6.4\e{+10}$ \\
Ra & 800  & 70.0 & $4.3\e{+10}$ \\
Sr & 800  & 67.5 & $8.4\e{+8}$ \\
Ba & 800  & 72.5 & $4.6\e{+8}$ \\
Mg & 800  & 55.0 & $6.3\e{+6}$ \\
Ca & 800  & 65.0 & $3.8\e{+6}$ \\
Be & 800  & 55.0 & $1.0\e{-7}$ 
\end{tabular}
\end{table}

\section{E1-M1 Optical Clock Stability}
\label{mono2}

The optimal clock stability, $\mathcal{S}$, can be expressed in terms of experimentally controlled parameters. Stability scales with the inverse square root of the effective rate of detected clock excitations, $\dot{N_D}$ (\ref{eq:ND}), and linearly with the linewidth, $\Delta\nu$

\begin{equation}
\mathcal{S} = \frac{\Delta\nu}{\nu_L}\sqrt{\frac{1}{\dot{N_D}\T}}.
\label{eq:stab2}
\end{equation}

\noindent The goal of an optical frequency standard is to minimize $\mathcal{S}$. This section focuses on the broadening mechanisms that constitute the linewidth of this system. An optical frequency standard ultimately measures a resonant laser frequency $\nu_L$. Resonance will be inevitably offset from the natural transition frequency, $\nu_0$, by shifts due to environmental interaction 

\begin{equation}
\nu_L = \nu_0+\nu_B.
\label{eq:bias}
\end{equation}

\noindent The natural transition frequency is equivalent among all like atoms. The biased resonance is offset by $\nu_B$ due to environmental perturbation of the atoms. This section also discusses the individual environmental shifts listed in Table \ref{dNu}. The total magnitude of $\nu_B$ can be reduced or eliminated by reducing and offsetting these experimental shifts. The individual broadening and shift mechanisms for the optimal Hg clock are discussed explicitly and shown in Table \ref{HbBroad}.\\

\begin{table}[h]%
\caption{The mechanisms that contribute to broadening $\Delta\nu$ and a bias frequency shift $\nu_B$ in a hot E1-M1 clock are listed. The broadening due to the light shift depends on the instability in laser intensity, $\sigma_{\bar{I}}$. Broadening due to black-body radiation will occur for temperature instability, $\sigma_{\bar{T}}$, but will be a much smaller effect than light shift broadening. }
\label{dNu}\centering %
\begin{tabular}{ lcc }
Mechanism &$\Delta\nu$ & $\nu_B$\\
\hline
Natural  & $\Delta \nu_{nat}$ &0 \T \\
Transit & $\Delta \nu_{TT} $ & 0 \\
Laserline & $\Delta \nu_{LL}$ & 0 \\
Doppler (1st-order)  & $\Delta \nu_{D1}$ & 0\\
Doppler (2nd-order)  & $\Delta \nu_{D2}$ &  $\nu_{D2}$ \\
Light Shift  & $\Delta\nu_{LS}(\sigma_{\bar{I}})$ & $\nu_{LS}$\\
Blackbody Radiation  &  $\Delta \nu_{BB}(\sigma_{\bar{T}})$ & $\nu_{BB}$\\
Collision  &  $\Delta\nu_{C}$ & $\nu_{C}$
\end{tabular}
\end{table}

\subsection{Broadening and Shift Mechanisms}
\subsubsection{Transit Broadening}
The dominant broadening mechanism in a hot vapor cell is the transit-time broadening, $\Delta\nu_{_{TT}}=1/\bar{t}$, which is introduced by the brief interaction time of the fast atoms through the narrow laser beam.  As such, the optimal beam radius $\omega_0$ to minimize $\mathcal{S}$ is large. Transit broadening scales with the dynamic parameters $\omega_0$ and $T$ as:

\begin{eqnarray}
\Delta\nu_{_{TT}}(T,\omega_0) &\propto& \frac{\sqrt{T}}{\omega_0}
\label{eq:Bias}
\end{eqnarray}

\noindent  For a hot Hg clock with minimal $\mathcal{S}$, transit broadening is the largest broadening mechanism. We calculate a mean interrogation time of $5 \mu s$, which introduces a Fourier uncertainty of $\Delta\nu_{TT}=0.2$ MHz. 
\subsubsection{Natural Width}

The clock state has highly suppressed relaxation channels, leading to long lifetimes and intrinsically narrow linewidths. Fermionic isotopes of Hg have estimated  natural linewidths, $\Delta \nu_{nat}$, of 0.5-0.7 Hz and the Bosonic isotopes have indefinite lifetimes \cite{Mishra2001}. These narrow widths are much smaller than the Fourier-limited width of the hot E1-M1 scheme imposed by the transit broadening.

\subsubsection{Light Shift and Broadening}

\label{LSStab}

The high levels of laser intensity required for non-zero transition probabilities $P_{^3P_0}$ (\ref{eq:probclock}) will introduce a.c. Stark shifts to the clock and ground state of the atoms. This will create a systematic bias in the fundamental frequency of the clock that scales with the intensity of the laser. Instability in laser intensity, $\sigma_{\bar{I}}$, will appear as a broadening to the system and uncertainty in the absolute intensity will manifest as an unknown fundamental bias. The dynamic dipole polarizability difference between the ${}^1S_0$ and ${}^3P_0$ levels of Hg at $531$ nm is 21 a.u. \cite{*[{Thanks to Guangfu Wang for provided the exact value at 531 nm.  }][{}] Ye2008}. The absolute shift scales with intensity as $2.25$ kHz/$\frac{W}{mm^2}$.

\subsubsection{Doppler Broadening}

First-order and second-order Doppler broadening both increase the linewidth of the frequency standard. The temperature dependence of first-order Doppler broadening is described by (\ref{eq:1D}). For misalignment less than 0.1 milliradians, the first-order Doppler broadening will contribute a maximum of 44 kHz of line broadening in Hg at 380 K. 

The temperature dependence of the second-order Doppler shift is described by: 
\begin{eqnarray}
 \nu_{D2}(T) = -\nu\frac{\bar{v}(T)^2}{2c^2}
\label{eq:2D}
\end{eqnarray}
 and is linear in temperature. The width of the velocity distribution is similar to the mean velocity in this range, so we assume $\Delta\nu_{D2} \approx |\nu_{D2}|$.  The hot Hg E1-M1 clock operates optimally at \OHgT , which introduces second-order Doppler broadening of 90 Hz. 

\subsubsection{Laser-Line Broadening}

The laser linewidth is the bottleneck through which all narrow linewidth information passes to the oscillation counter in an optical frequency standard. Even if a broad linewidth laser is resonant with and centered on a narrow transition, the ultimate frequency measurement will not register frequency features narrower than the those of the laser. It is therefore important for the laser linewidth to be narrower than the effective linewidth of the atom. For Hg, a sub-kHz linewidth specification will make a negligible contribution to the overall broadening due to the large size of transit broadening.

\subsubsection{Black-body Radiation Shift and Broadening}

The estimated black-body radiation shifts and uncertainties have been cataloged elsewhere \cite{Mitroy2010}. At \OHgT$\,$ the shift is $-1.63$ Hz. Inaccuracy and instability, $\sigma_{\bar{T}}$, in operating temperature can introduce an unknown systematic shift and broadening. This is typically much smaller than other broadening and shift features due to the overall small size of the radiation shift.

\subsubsection{Collision Shift and Broadening }
\label{collshift}
 The specific collision phase shift introduced by colliding $^1S_0$ and $^3P_0$ Hg atoms is unknown and will need to be measured experimentally. All the group II atoms, with the exception of Hg, have less than 0.1\% probability of experiencing a collision in the excitation region. At the optimal operating temperature for the hot Hg clock, the collision frequency ($\dot{C} = \rho\sigma\bar{v}$) is 1.5 $\e{5}$ s$^{-1}$. We capture the harm collisions will have on the stability in the worst-case by treating collision as a signal loss channel. 

Collision broadening of the clock level transition has been observed in ultracold Sr \cite{Lisdat2009}. If we extrapolate that result to Sr number densities in a vapor cell at 800 K, we can estimate collision broadening of Sr to be $\Delta\nu_{C} = 31$ kHz which compares to the estimated first-order Doppler broadening at this temperature of $\Delta\nu_{D1} = 32$ kHz. This suggests that in general collision broadening will be less harmful than the collision loss approximation we've made in this paper.




\begin{table}[h]%
\caption{Broadening and shift budget for monochromatic, E1-M1 Hg clock at the temperature (\OHgT) and laser beam radius (0.6 mm) where the minimum \mcS is found.}
\label{HbBroad}\centering %
\begin{tabular}{ lcc }
Mechanism & Broadening [Hz] & Shift [Hz]\\
\hline
Transit & 2$\e{5}$ & 0\T \\
    Doppler 1st & 4$\e{4}$ & 0 \\
    Doppler 2nd & 90 & -90 \\    
    BBR & $10^{-16}$ & -1.7\\ 
Stark & 314 & 1.6$\e{4}$ \\
Natural & 0.45 & 0 \\
   \hline
   $\Delta\nu$ & 2$\e{5}$  &\T \\
  $\nu_B$ & &1.6$\e{4}$   \\
\end{tabular}
\end{table}

\subsection{Optimal Clock Stability}

Optimal stability for each group II type atom can be achieved through selection of laser beam radius and vapor cell temperature. The ideal laser beam radius is defined by the crossover point between time-limited $\Delta\nu_{TT}$ (\ref{eq:Bias}) and velocity-limited $\Delta\nu_{D1}$ (\ref{eq:1D}) broadening regimes. This crossover occurs in a detection-limited volume, so we use the $L_o$ scaling of the effective clock excitation rate (\ref{eq:NTw}). We characterize the scaling behavior of these broadening regimes with respect to laser beam radius $\omega_0$ and temperature $T$ to determine the minimum $\mathcal{S}$ (\ref{eq:stab2}):

\begin{eqnarray}
\mathcal{S} & \propto & { \Delta\nu_{TT} \brace \Delta\nu_{D1}} \times \sqrt{\frac{1}{\dot{N}_{^3P_0} \T}}\\
\mathcal{S}(T,\omega_0)  & \propto &   {\sfrac{\sqrt{T}}{\omega_0} \brace \sqrt{T}}  \times  \sqrt{\frac{\omega_0 \sqrt{T}}{e^{\frac{-10^4}{T}}} } \nonumber \\
 & \propto &  \sqrt{\frac{T\sqrt{T}}{e^{\frac{-10^4}{T}}}{\frac{1}{\omega_0} \brace{\omega_0}} }.
\label{eq:StabScale}
\end{eqnarray}

\noindent Curly brackets no longer denote the geometry limits of (\ref{eq:cbgeo}), instead they are the time- and velocity-limited behavior ${ \Delta\nu_{TT} \brace \Delta\nu_{D1}}$. As with maximum excitation rates $\dot{N}_{^3P_0}$, optimally small \mcS  favors high temperature $T$ (\ref{eq:NTw}). Unlike maximum $\dot{N}_{^3P_0}$, optimal \mcS favors a larger laser beam radius $\omega_0$ than one that matches the Rayleigh range to the detection-limit. This is due to \mcS's sensitivity to transit-time broadening $\Delta\nu_{TT}$. 

\begin{figure}[htp]
\includegraphics[width=3in]{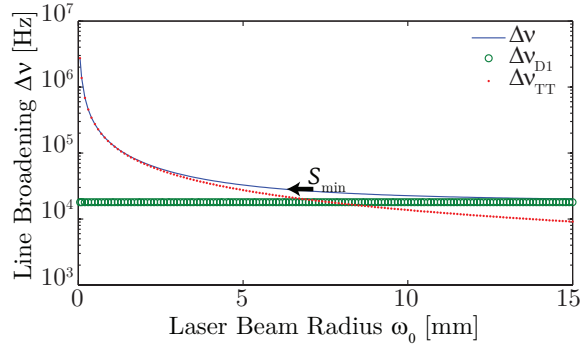}
\caption{\textbf{Asymptotic Behavior of $\boldsymbol{\Delta\nu}$ in Ra} Calculation results for laser-beam radius $\omega_0$ dependence of the transit broadening $\Delta\nu_{TT}$, first-order Doppler broadening $\Delta\nu_{D1}$, and total broadening $\Delta\nu$ for Ra at vapor cell temperature 800 K .  
\label{F:saRa}}
\end{figure}

Figure \ref{F:saRa} shows the ideal laser-beam radius $\omega_0$ for minimum \mcS is closely related to the crossover point between time-limited $\Delta\nu_{TT}$ (\ref{eq:Bias}) and velocity-limited $\Delta\nu_{D1}$ (\ref{eq:1D}) broadening regimes. We show the behavior in Ra because its stability behavior with respect to laser beam radius is representative of all the group II type atoms except Hg. The competing advantage of increased excitation rate $\dot{N}_{^3P_0}$ at smaller $\omega_0$ in this detection-limited regime pushes the final \mcS radius slightly shorter than the crossover length.

\begin{figure}[htp]
\begin{center}
\mcS  [$\sqrt{\mathrm{Hz}^{-1}}$] 
\includegraphics[width=3.33in]{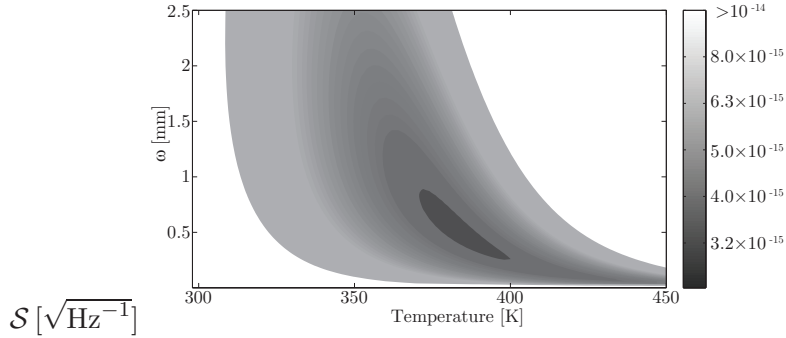}
\caption{\textbf{Experimental Parameters for Hg Stability}
The stability \mcS (\ref{eq:stab2}) of a hot Hg E1-M1 clock is plotted for laser-beam radius $\omega_0$ and vapor cell temperature $T$. \mcS$_{\text{min}}$= \OHgS  is found at $\omega_0=0.6$ mm and $T=380$ K. 
\label{HgStab}}
\end{center}
\end{figure}

Figure \ref{HgStab} shows the specific behavior of Hg's stability with respect to laser beam radius and temperature, where the darkest region has the lowest stability. The collision-free scaling of the stability favors maximum vapor cell temperture (\ref{eq:StabScale}). Hg is affected by collisions and thus is optimized at a lower temperature $T$ and smaller beam radius $\omega_0$ than the other group II type atoms.  \

The reduction of transit broadening is important to reduce \mcS in a hot E1-M1 clock, and insensitivity to transit broadening among atomic species scales with mass. Table \ref{D1} shows relative first-order Doppler broadening for the group II type atoms. Table \ref{Stb} shows the optimal temperature and beam size for each element to produce minimum \mcS in a monochromatic scheme calculated with the experimental assumptions in Table \ref{EAss}.

\begin{table}[htp]%
\caption{The vapor cell temperature $T$ and laser beam radius $\omega_0$ to achieve minimum $\mathcal{S}$ for each group II type atom is listed. Stability \mcS is defined in (\ref{eq:stab}) and characterizes how quickly a frequency standard can achieve a chosen absolute accuracy.} 
\label{Stb}\centering %
\begin{tabular}{ c c c l}
Atom & T [K]& $\omega_0 [mm]$ & $\mathcal{S}$ [$\sqrt{\mathrm{Hz}^{-1}}$]  \\
\hline
Hg & 380  & 0.6 & $1.6\e{-15}$\T \\
Ra & 800  & 6.0 & $3.5\e{-15}$ \\
Yb & 800  & 2.9 & $4.2\e{-15}$ \\
Sr & 800  & 5.0 & $4.2\e{-14}$ \\
Ba & 800  & 7.0 & $4.3\e{-14}$ \\
Ca & 800  & 6.0 & $8.3\e{-13}$ \\
Mg & 800  & 1.9 & $1.2\e{-12}$ \\
Be & 800  & 4.0 & $9.6\e{-6}$ 
\end{tabular}
\end{table}

\begin{table}[htp]%
\caption{Hg parameters at the $\mathcal{S}_{\text{min}}$.}
\label{T:StHg}\centering %
\begin{tabular}{ l l r  }
Parameter & & Value\\
\hline
\mcS$_{\text{min}}$ & stability (\ref{eq:stab2}) & 3.1\e{-15} $\sqrt{\text{Hz}^{-1}}$\T\\
$\omega_0$ & laser-beam radius & 0.54 mm\\
$T$ & vapor cell temperature & 382 K\\
$\rho$ & number density & 10$^{22}$ m$^{-3}$\\
$P_{^3P_0}$ &excitation probability (\ref{eq:probclock})& 7.5\e{-8}\\
$\dot{N}_{tot}$ &atom interrogation rate (\ref{eq:ntot})& 3.8\e{+19} s$^{-1}$\\
$\Delta\nu$ &effective linewidth - Table \ref{dNu} & 190 kHz\\
$\Omega_{R 2\gamma}$ &two-photon Rabi frequency (\ref{eq:2gRates}) & 100 Hz\\
$\bar{t}$ &interrogation time (\ref{eq:tB}) & 5.38 $\mu$s\\
$\dot{C}$ & collision frequency & 1.5 $\e{5}$ s$^{-1}$
\end{tabular}
\end{table}

In all, we have shown hot clock stabilities that are competitive with current cold standards. They can be achieved with a single excitation laser driving a degenerate two-photon E1-M1 transition and using commercially available systems. This clock scheme could fill the currently empty niche of a portable optical frequency standard.

\section{Bichromatic E1-M1 Clock\newline Light Shift Elimination}
\label{bich}

Introducing a second excitation laser frequency in the E1-M1 clock scheme creates an opportunity to eliminate the light shift completely. Current lattice clocks use a magic wavelength for atom trapping to balance the dynamic light shift of the ground and clock levels due to the lattice laser, but they remain subject to a light shift from the excitation beam \cite{Poli2008}. Methods to mitigate the impact of the excitation light shift on an optical clock transition have been explored directly \cite{Zanon-Willette2006} or by extension \cite{Huntemann2012a,Yudin2010a}.\

Complete elimination of the light shift during excitation is possible in an E1-M1 clock with dual frequencies. Wavelength can be selected to resonate with the transition and laser intensity can be chosen to offset light shift differences perfectly. The dynamic dipole polarizabilities of neutral Hg have been determined sufficiently across optical frequencies \cite{Ye2008} to predict the magic wavelength pair that will fully eliminate the light shift during optical excitation of the clock transition for a set of lasers with equal intensity. The wavelengths and dynamic dipole polarizability differences $\Delta \alpha(\lambda)$ of this magic pair solution are listed in Table \ref{MLs}.

\begin{table}[h]%
\caption{A magic wavelength pair for neutral Hg is shown. This pair of wavelengths $\lambda$ eliminate the light shift when the electric field of each laser is equal in magnitude. The dynamic dipole polarizability of the ground state $\alpha_{^1S_0}(\lambda)$ and clock state $\alpha_{^3P_0}(\lambda)$ are listed along with the dynamic dipole polarizability difference $\Delta \alpha(\lambda)$. These values were calculated in \cite{Ye2008}. } 
\label{MLs}\centering %
\begin{tabular}{ c c c c }
$\lambda [nm]$ & $  \alpha_{^1S_0}(\lambda)$ [a.u.] & $  \alpha_{^3P_0}(\lambda)$ [a.u.]& $\Delta \alpha(\lambda)$  \\
\hline
376 & 39 & 10 & $-29$ \T \\
905 & 32 & 61 & $+29$ 
\end{tabular}
\end{table}

The bichromatic scheme enjoys a smaller intermediate detuning $\Delta$ from the ${^3P_1}$ level than the monochromatic scheme. Reduction of $\Delta$ leads to a relative increase in the resonant two-photon Rabi frequency (\ref{eq:2gRates}). The specific two-photon Rabi frequency of the magic pair solution in Table \ref{MLs} will enjoy a factor of 4.9 increase compared with the monochromatic solution of the hot Hg clock. Rate estimates are based on adiabatic elimination of the intermediate level and this approximation may begin to degrade \cite{Brion2007} with reduced intermediate detuning $\Delta$. Implementation of a bichromatic method to eliminate a light shift will elevate first-order Doppler broadening to unacceptable levels in a hot clock and thus requires ultracold atoms or atoms with a uniform velocity direction (for example in a thermal beam). Conveniently for an ultracold scheme, either excitation beam can serve as a red-detuned optical dipole trap cycling on the 265-nm ${^1S_0} \xleftrightarrow{\text{E1}} {^3P_1}$ transition. The trap laser can combine with its magic pair from another laser to complete a bichromatic E1-M1 excitation. The lasers used to excite the transition introduce offsetting light shifts to completely negate the light shift bias $\nu_{LS}$.

\section{conclusion}

A hot E1-M1 Hg vapor cell can achieve small \mcS$\,$ comparable to the current minimum found in the Sr lattice clock \cite{Bloom2014}. Table \ref{shStand} summarizes this along with other state-of-the-art frequency standards. We include the Rb chip \cite{Knappe2005} to allow comparison with a portable microwave standard. 

\begin{table}%
\caption{The stability \mcS and accuracy $\sigma_\nu$ of current optical frequency standards. The fundamental frequency $\nu$, linewidth $\delta\nu$, and detected atom number $N$ are also listed for these systems. Brackets in this table indicate order of magnitude, where a magnitude $n_1\e{n_2}$ is reported as $n_1[n_2]$. \newline
(\textbf{bold} denotes predicted values)\newline
$fn$ - fountain, $cp$ - chip, $ion$ - ion, $lt$ - lattice,\newline
$vp$ - hot vapor cell}
\label{shStand}\centering %
\begin{tabular}{ lcccrr }
Atom & $\nu$ [Hz] & $\delta \nu$ [Hz] & $N$ [\#] &  $\mathcal{S}$ & $\sigma_\nu$ \\ 
\hline
Rb$_{cp}$\cite{Knappe2005}  & $6.8[9]$ &  &  & $6[-11]$ & 6[-12] \\
Cs$_{fn}$\cite{Jefferts2007}   & $9.1[9]  $ & & 1[7] & 2[-13] & 5[-16] \\
Hg$_{vp}$  & $5.6[14]$ & $\boldsymbol{2[5]}$ & $\boldsymbol{2.3[21]}$  & $\boldsymbol{1.6[-15]}$ & \\
 Al$^+_{ion}$\cite{Chou2011,Chou2010}  & $1.1[15]$ & $7$ & $1$  & $3.7[-16]$ &  $8.6[-18]$\\
Yb$_{lt}$ \cite{Hinkley2013}  & $5.2[14]$ & $6$ & $5[3]$  & $3.2[-16]$ & $1.6[-18]$\\
Sr$_{lt}$ \cite{Bloom2014}  & $4.3[14]$ & $6-50$ & $2[3]$  & $3.1[-16]$ & $6.4[-18]$\\
\end{tabular}
\end{table}

The prospects of a hot E1-M1 clock as a portable frequency standard are compelling. The monochromatic Hg scheme enjoys a large number of addressed atoms to improve its precision statistics. It suffers from a large absolute frequency shift due to the intensity of the incident beam, but the accuracy of this standard can be preserved with precise laser power metrology.

A frequency standard unbiased by the light shift due to excitation is for the first time possible with a bichromatic E1-M1 scheme. A two-photon excitation scheme may enable current cold systems to completely eliminate systematic frequency bias. Such an absolutely accurate frequency standard can be used to make distributed and temporal measurements of local gravitational redshift and variation of the fine structure constant $\alpha$.

We have characterized the E1-M1 clock. We have shown how the clock-level excitation rate for each group II type atom depends on the experimentally-controllable parameters vapor cell temperature and laser beam radius. We have calculated that neutral Hg is the optimal atomic system for a hot optical clock, where its main advantage is a relatively high number density compared to the other group II type atoms. With the conservatively selected experimental parameters assumed in this paper, we calculate the stability for a hot Hg clock could be as low as \OHgS. This stability is competitive with other optical frequency standards while offering the portability of a vapor cell.

\bibliography{doc/EAtBib}

\end{document}